\documentclass[conference]{IEEEtran}
\IEEEoverridecommandlockouts
\makeatletter
\def\endthebibliography{%
  \def\@noitemerr{\@latex@warning{Empty `thebibliography' environment}}%
  \endlist
}
\usepackage{cite}
\usepackage{amsmath,amssymb,amsfonts}
\usepackage{algorithmic}
\usepackage{graphicx}
\usepackage{textcomp}
\usepackage{xcolor}
\usepackage{verbatim}
\usepackage{array}
\usepackage{url}
\usepackage{comment}
\usepackage{esdiff}
\usepackage{multirow}
\usepackage{multicol}
\usepackage{svg}
\usepackage[symbol]{footmisc}
\def\BibTeX{{\rm B\kern-.05em{\sc i\kern-.025em b}\kern-.08em
    T\kern-.1667em\lower.7ex\hbox{E}\kern-.125emX}}
\usepackage{caption}
\usepackage{float}
\usepackage{subcaption}
\renewcommand{\baselinestretch}{0.97}

\begin{document}

\title{ViTA: A Vision Transformer Inference Accelerator for Edge Applications}

\author{\IEEEauthorblockN{Shashank Nag\IEEEauthorrefmark{1}, Gourav Datta\IEEEauthorrefmark{2}, Souvik Kundu\IEEEauthorrefmark{3}, Nitin Chandrachoodan\IEEEauthorrefmark{1}, Peter A. Beerel\IEEEauthorrefmark{2}}
\IEEEauthorblockA{\textit{\IEEEauthorrefmark{1} Department of Electrical Engineering, Indian Institute of Technology Madras, India}}
\IEEEauthorblockA{\textit{\IEEEauthorrefmark{2} Ming Hsieh Department of Electrical and Computer Engineering, University of Southern California, USA}}
\IEEEauthorblockA{\textit{\IEEEauthorrefmark{3} Intel Labs, San Diego, USA}}
\textit{\{shashank@smail, nitin@ee\}.iitm.ac.in,}
\textit{\{gdatta, pabeerel\}@usc.edu,}
\textit{souvikk.kundu@intel.com}
\thanks{This work was supported in part by the Indo-US Science \& Technology Forum and Viterbi School of Engineering, University of Southern California, under the IUSSTF-Viterbi India Program 2022.}
}

\maketitle

\begin{abstract}
Vision Transformer models, such as ViT, Swin Transformer, and Transformer-in-Transformer, have recently gained significant traction in computer vision tasks due to their ability to capture the global relation between features which leads to superior performance.
However, they are compute-heavy and difficult to deploy in resource-constrained edge devices. Existing hardware accelerators, including those for the closely-related BERT transformer models, do not target highly resource-constrained environments. In this paper, we address this gap and propose ViTA - a configurable hardware accelerator for inference of vision transformer models, targeting resource-constrained edge computing devices and avoiding repeated off-chip memory accesses. We employ a head-level pipeline and inter-layer MLP optimizations, and can support several commonly used vision transformer models with changes solely in our control logic. We achieve nearly 90\% hardware utilization efficiency on most vision transformer models, report a power of 0.88W when synthesised with a clock of 150 MHz, and get reasonable frame rates - all of which makes ViTA suitable for edge applications. 
\end{abstract}

\begin{IEEEkeywords}
Vision Transformer, Swin Transformer, Hardware Accelerator, Computer Vision, Edge Computing, FPGA
\end{IEEEkeywords}
\section{Introduction}\label{intro}
The success of transformer models for NLP applications has led to self-attention-based models being applied to computer vision tasks. Although the initial works in this direction lacked scalability, subsequent works such as Vision Transformers (ViT) \cite{vit}, Swin Transformers \cite{swin}, TNT \cite{tnt}, and Data-efficient image Transformers (DeiT) \cite{deit} that adopted model architectures similar to the NLP-based Transformer \cite{vaswani}, replacing word tokens with image patches, have yielded state-of-the-art (SOTA) results.

For some applications, such as autonomous driving and drone navigation, computer vision tasks have demanded real-time implementations on the edge. This has led to the development of energy-efficient hardware accelerators such as Eyeriss~\cite{eyeriss,eyeriss-v2}, ShiDianNao \cite{shidiannao}, \cite{fpga-cnn} for inferences of traditional CNN-based models \cite{kundu2020pre}. With vision transformer models outperforming the conventional CNN-based models, exploring energy-efficient hardware accelerators for ViTs, which could be implemented at the edge, is similarly important. 
%

FPGAs for edge applications are useful because they provide flexibility in computation while still enabling very low latencies.  For example, the Zynq Multiprocessor System-on-Chip (MPSoC) ZC7020~\cite{li-paap21} and low-end Ultrascale MPSoC ZU3EG~\cite{suh-icfpt21,yang-fpga19} are some FPGAs that can be used in applications such as drones and similar low-energy vision oriented tasks. The defining characteristics of such platforms are limited parallelism (100s of DSP units rather than 1000s) and limited on-chip memory (100s of KB of Block RAM (BRAM) rather than MB). Naturally, this means any application targeting such platforms should focus on using the limited available parallelism and minimizing the amount of data transfer to and from off-chip memory.

In this work, we propose ViTA - a hardware accelerator architecture and an efficient dataflow that supports several popular vision transformer models, targeting such resource-constrained FPGA devices. We evaluate the performance of ViTA for these models with commonly used configurations in terms of both hardware utilization efficiency and throughput. 

The remaining of this paper is organized as follows. We provide a primer of the different vision transformer models, the computations involved therein, and existing related hardware accelerators in Section \ref{sec:background}. We propose ViTA - an architecture suitable for edge devices and an efficient dataflow within typical memory bandwidth constraints in Section \ref{sec:architecture}. We evaluate our design for different model architectures and configurations in Section \ref{sec:analysis}. Section \ref{sec:conclusions} concludes the paper.

\section{Preliminaries and Related Work}\label{sec:background}

\begin{figure}
    \centering
    \begin{subfigure}{2.7cm}
      \includegraphics[width = 2.7cm]{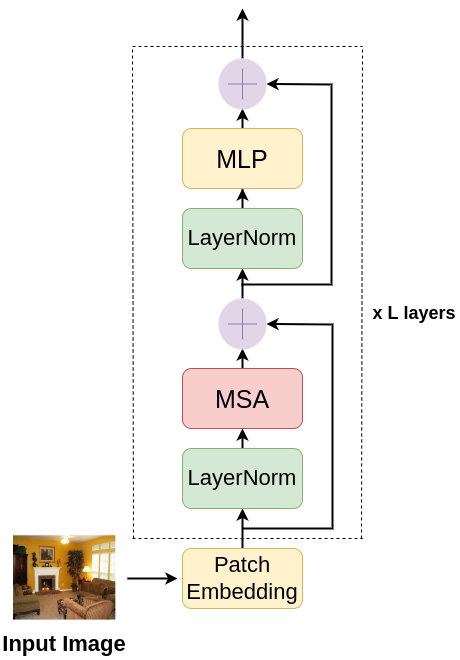} 
      \caption{ViT / DeiT}
      \label{fig:vit}
    \end{subfigure}
    \begin{subfigure}{5.5cm}
      \includegraphics[width = 5.5cm]{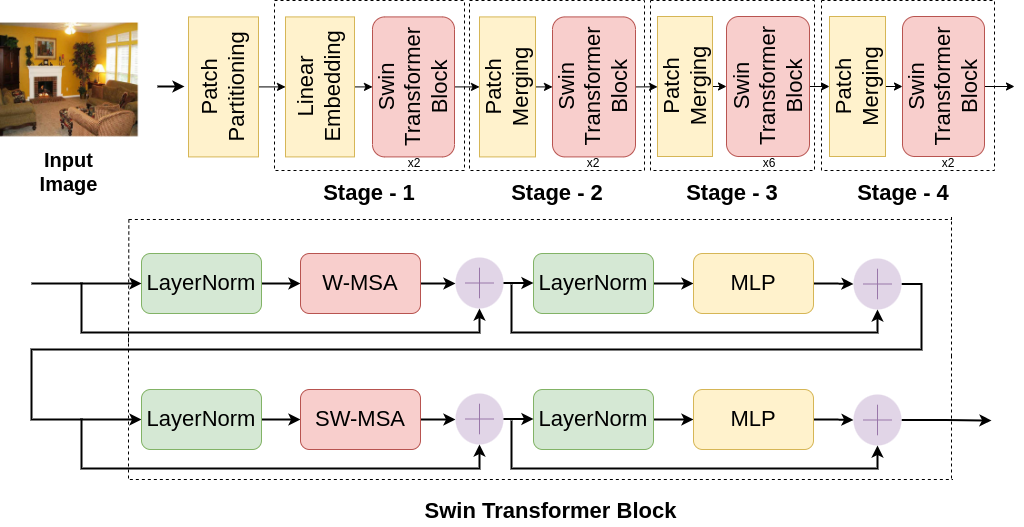} 
      \caption{Swin Transformer - two successive Swin Transformer blocks have a windowed and shifted windowed MSA}
      \label{fig:swin}
    \end{subfigure}
    \caption{Vision Transformers under consideration. The Swin Transformer has the same set of blocks as ViT, but the MSA is applied on disjoint windows on the image, and the patch merging blocks scales the image dimensions down in successive stages.}
    \label{fig:visiontrans_models}
    \vspace{-0.5cm}
\end{figure}

The Vision Transformer (ViT) \cite{vit} is one of the first proposed transformer-based models for vision tasks, and is similar to the encoder stack of the BERT transformer \cite{vaswani}. In ViT, the input image is split into patches of 16$\times$16 pixels, which constitute a linear sequence of tokens, similar to  words in the case of BERT. Its success motivated the development of other transformer-based models, such as the Swin Transformer \cite{swin}, Data-efficient image Transformers \cite{deit}, and Transformer-in-Transformer \cite{tnt}. The key operation of all these models is the Multi-head Self Attention (MSA), applied on the sequence of image patches, followed by fully connected layers. The variation among these models is related to how these attention blocks are applied to the input patches, and the model parameters such as latent space dimensions, patch sizes, and the number of heads, for each of their variants. Figure \ref{fig:visiontrans_models} illustrates some of these model architectures.

There have been a few works on hardware accelerators for vision transformers \cite{vit-hw}\cite{auto-vit}\cite{vaqf}. Wang et al. \cite{vit-hw} targets an application-specific integrated circuit (ASIC), but does not consider off-chip data movement optimizations. Li et al. \cite{auto-vit} and Sun et al. \cite{vaqf} target a large FPGA device, which enables storing all the weights and activations on-chip, thereby allowing a larger design space for exploration. However, edge computing devices, such as the Zynq ZC7020 MPSoC, pose more severe design constraints, owing to the lower on-chip computational resources, BRAM memory, and bandwidth for off-chip access. In particular, off-chip memory accesses typically involve high energy and should be minimized.

As could be seen from the model architectures, the vision transformer models are quite similar to the BERT transformer \cite{vaswani}. There have been prior works on hardware accelerators targeting different stages of the BERT transformer. Liu et al. \cite{quantised-bert} proposed an accelerator for a quantized BERT model, and Lu et al. \cite{Lu2020} proposed an accelerator design for the MSA and feed-forward blocks, with dedicated computation units for Softmax and LayerNorm. Li et al. \cite{Li2020} proposed FTRANS, an acceleration framework for large scale language representations. However, since NLP applications are typically not deployed at the edge, these target higher-end FPGA devices, and do not consider data movement from off-chip storage devices. ReTransformer \cite{retransformer} proposed a Re-RAM based \cite{fayyazi2019csrram} in-memory computation engine with granular pipeline to accelerate the self-attention stage. In-memory computations typically pose additional limitations in terms of correctly aligning the results into the in-memory compute unit, which can be avoided in FPGA-based digital designs. However, the granular pipeline aspect of this work could potentially reduce the intermediate memory requirements while ensuring high efficiency, and we could explore this aspect in our design. 

\section{Proposed Architecture}\label{sec:architecture}
\subsection{Model Architecture}
We base our design on the ViT-B/16 model \cite{vit}, and in Section \ref{sec:analysis} show how it can be extended to other vision transformers. The ViT-B/16 model considers a patch size $(P\times{P})$ where $P{=}16$, having $L{=}12$ layers, a latent vector dimension $(D)$ of 768, $k{=}12$ heads, a head level latent vector dimension $(D_h = D/k)$ of 64, and an MLP hidden layer dimension $(M)$ of 3072 \cite{vit}.  We consider an image dimension $(H\times{W})$ of $256\times256$, thus giving the sequence length $(N)$ as $(H.W)/P^2$ = 256.

We follow a post-training quantization approach, with all the weights and activations quantized to int8 representations for inference. We observe that when evaluated on ViT, this results in almost no degradation (${<}0.04\%$) in the top-1 test accuracy on the ImageNet \cite{imagenet_cvpr09} dataset. 

\subsection{Hardware Architecture and Dataflow} \label{ss:had}

We target the Zynq ZC7020 MPSoC (an edge FPGA) for our design. As noted previously, targeting edge computing devices poses multiple constraints. Table \ref{tab:memreq} shows the memory requirement for the ViT-B model, which is beyond the typical on-chip memory capacity of edge FPGA devices, indicated in Table \ref{tab:resavlbl}. This necessitates most of the data to be stored in off-chip DRAMs, and brought into the on-chip BRAM cells when required, while minimizing back-and-forth data movement. 

\begin{table}[h]
\centering
\begin{tabular}{l|l|l|l|l|l}
\hfil Input/Weights & \hfil Input & \hfil $W^Q$ & \hfil $W^K$ & \hfil $W^V$ & \hfil MSA Weights \\
\hline
\hfil Memory (in KB) & \hfil 192 & \hfil 576 & \hfil 576 & \hfil 576 & \hfil 576 \\
\end{tabular}
\caption{Memory requirements for the input activations and weights for the ViT-B/16 model. The memory requirements for intermediate results are significantly higher.}
\label{tab:memreq}
    \vspace{-0.5cm}
\end{table}

\begin{table}[h]
\centering
\begin{tabular}{l | l | l}
 & Zynq ZC7020 MPSoC & Zynq ZCU102 \\
\hline
\hfil LUT6  & \hfil 53200 & \hfil 274080\\
\hfil DSP slices  & \hfil 220 & \hfil 2520\\
\hfil BRAM & \hfil 630 KB & \hfil 4 MB\\
\end{tabular}
\caption{Resources available on the Zynq ZC7020 (embedded) MPSoC vs. the Zynq ZCU102 (high-end) MPSoC}
\label{tab:resavlbl}
\vspace{-0.2cm}
\end{table}

Most of the computations in inference of vision transformers are matrix multiplications. Schemes such as input-stationary, weight-stationary, or block multiplication can be adopted to meet on-chip memory constraints. We note that in all the vision transformer architectures in Fig. \ref{fig:visiontrans_models}, multiple layers of self attention and MLP blocks are stacked one after the other. While the weights are different for each of these layers, the input activations for a layer are passed from the previous layer. This suggests that an input stationary scheme, with the weights loaded to the chip on-demand would be ideal. Once the input activations are loaded, the results of each layer could be stored in the same on-chip location, and only the final results (after all the layers) need to be written back to off-chip memory. Given this arrangement, it would be ideal to hide the off-chip memory access latency involved in fetching the weights, by scheduling computations and weight fetching appropriately. Our proposed approach is to implement this schedule at a column level of the weight matrix. For each weight matrix, BRAM cells that can store two columns of weights are allocated. The access latency can be hidden by ensuring that while one column of the weights is being operated upon, the other column is fetched from off-chip.

\begin{table}[h]
\centering
\begin{tabular}{l | l | l | l | l}
\hfil Model & \hfil Image Dim. & \hfil MSA & \hfil MLP & \hfil Patch Merging \\
\hline
\hfil ViT-B/16 \cite{vit} & $256\times256\times3$ & \hfil 36.8\% & \hfil 63.2\% & \hfil -\\
\hfil ViT-B/16 \cite{vit} & \hfil $224\times224\times3$ & \hfil 36.1\% & \hfil 63.9\% & \hfil -\\
\hfil DeiT-S \cite{deit} & \hfil $224\times224\times3$ & \hfil 38.6\% & \hfil 61.4\% & \hfil -\\
\hfil DeiT-T \cite{deit} & \hfil $224\times224\times3$ & \hfil 43.1\% & \hfil 56.9\% & \hfil -\\
\hfil Swin-T \cite{swin}  & \hfil $224\times224\times3$ & \hfil 31.9\% & \hfil 63.8\% & \hfil 4.3\%\\
\end{tabular}
\caption{Number of MAC operations in the MSA, MLP, and Patch Merging layers as a fraction of the total number of MAC operations.
We ignore the Softmax, LayerNorm, and Residual Connection layers.}
\label{tab:comps}
    \vspace{-0.4cm}
\end{table}

\subsubsection{Architecture for MLP Layer}\label{sssec:mlp}
\begin{figure}
    \centering
    \begin{subfigure}{8cm}
      \centering
      \includegraphics[width = 8cm]{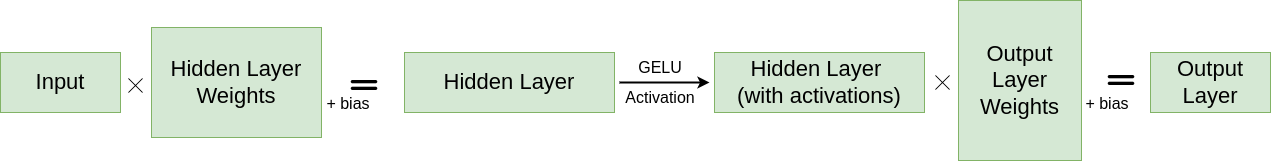} 
      \caption{MLP Block}
      \label{label}
    \end{subfigure}
    \begin{subfigure}{8cm}
      \centering
      \includegraphics[width = 7cm]{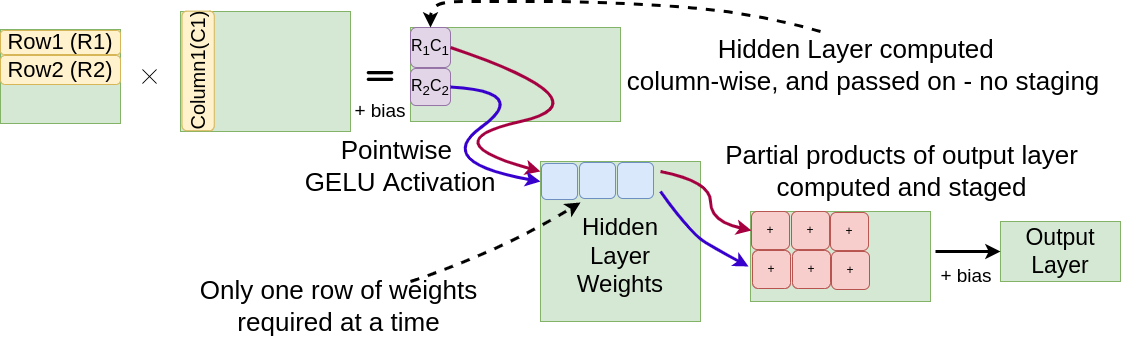} 
      \caption{Scheduling MLP operations in two stages}
      \label{label}
    \end{subfigure}
    \caption{Inter-layer optimization for MLP}
    \label{fig:mlp}
        \vspace{-0.4cm}
\end{figure}

As noted in Table \ref{tab:comps}, the majority of MAC computations are in the MLP layer, accounting for nearly 60\% of the total MAC operations across the different model variants. Thus, it is pertinent to accelerate this computation efficiently. 
The design for a simple feed-forward layer could be trivial, with the same weights and biases being applied on multiple input activations concurrently. However, we observe that the memory required to store the hidden layer, with dimensions of 256$\times$3072 for the ViT-B model, is beyond the total on-chip memory capacity of the target device. To avoid repeated DRAM accesses, we employ the inter-layer optimization technique proposed in \cite{inter_layer}, using two sets of MAC units. As elaborated in Fig. \ref{fig:mlp}, the hidden layer values computed in the first set of MAC units are broadcast to the second set of MAC units through the non-linear activation, to compute the partial products corresponding to the output layer. This leads to one of the key ideas that inspire our design. In order to process this in a pipelined fashion, we allocate resources such that the hidden layer value computations and the output layer partial product computations take approximately equal times. This ensures they can be pipelined with minimal stalls. This implies that we should dedicate an equal number of MAC units to compute the hidden and output layers. While the MACs dedicated to the hidden layer accumulate the results over multiple cycles, those dedicated to the output layer compute different partial products in each of those cycles.

\subsubsection{Architecture for MSA Layer}\label{sssec:msa}
For all the model architectures illustrated in Table \ref{tab:comps}, the self attention operation also accounts for about 30-40\% of the total computations. Consequently, we try to optimize the hardware accelerator design to suit these computations, while adhering to the above scheme for the MLP layer. The MSA layer involves the following computations \cite{kundu2021attentionlite}, where $z$ refers to the input to the layer, $i$ to the head number, $k$ to the total number of heads, and Q, K, V \& SA to the standard self-attention terminology:
\begin{equation}
    [Q|K|V]_i = z.[W^Q|W^K|W^V]_i  \quad ,   W^{Q/K/V} \in R^{D\textrm{x}D_h}
\end{equation}
\begin{equation}
    S_i(z) = \textrm{Softmax}(Q_iK^T_i/\sqrt{D_h})
\end{equation}
\begin{equation}
    SA_i(z) = S_i.V_i
\end{equation}
\begin{equation}
    MSA(z) = [SA_1(z),...,SA_k(z)].W^{msa} \quad , W^{msa} \in R^{D\textrm{x}D}
\end{equation}

where (.) refers to matrix multiplication. As illustrated, the MSA block involves computing self attention on each head, and finally concatenating the results. Although the operations on each head can be parallelized, the on-chip resource constraints dominate the design choice. Processing multiple heads implies that the computed values (say $Q$, $K$, and $V$) for all the heads will have to be staged until the next set of values ($Q.K^T$, $S$, and $SA$) are computed. Given the on-chip memory constraints, storing these intermediate matrices for all the heads is not feasible. Hence, to avoid unnecessary data movement to and from off-chip, we instead perform head-wise computations. As detailed in Sec \ref{sec:background}, there have been several works focused on accelerating the MSA block, such as the ReTransformer \cite{retransformer}, which proposes a row-level fine-grained pipeline for efficiently computing the self attention block using two PE engines. However, since the input matrix has been held stationary, the remaining available memory is not sufficient to hold the required weights for computing the $Q$, $K$, and $V$ for a head. To avoid repeated off-chip accesses, we, in contrast, store the weights one column at a time and proceed to the next column once all rows of input activations are multiplied with this column. As a consequence, we can only compute the $Q$, $K$, and $V$ matrices in a column-wise fashion, and hence the $QK^T$ and further computations for a given head cannot occur until both $Q$ and $K$ matrices are completely computed. To this effect, we propose a head-level coarse-grained pipeline with two dedicated sets of processing units handling these different computations, with near-equal latency.

\subsubsection{Overall Architecture and Optimal Configuration}
\begin{figure}
    \centering
    \begin{subfigure}{8cm}
      \centering
      \includegraphics[width = 7cm]{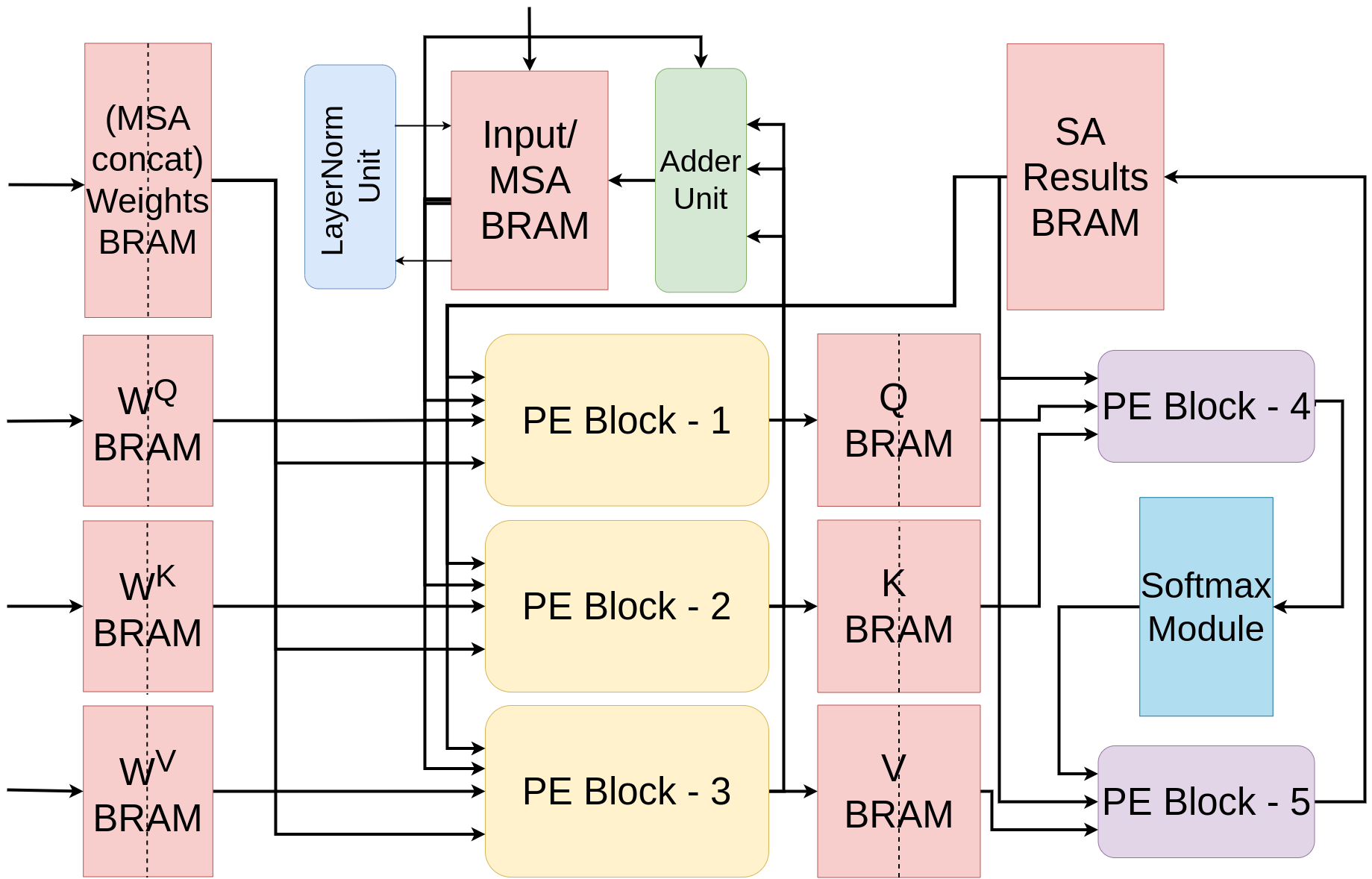} 
      \caption{Overall Hardware Accelerator Design}
      \label{label}
    \end{subfigure}
    \begin{subfigure}{8cm}
      \centering
      \includegraphics[width = 7cm]{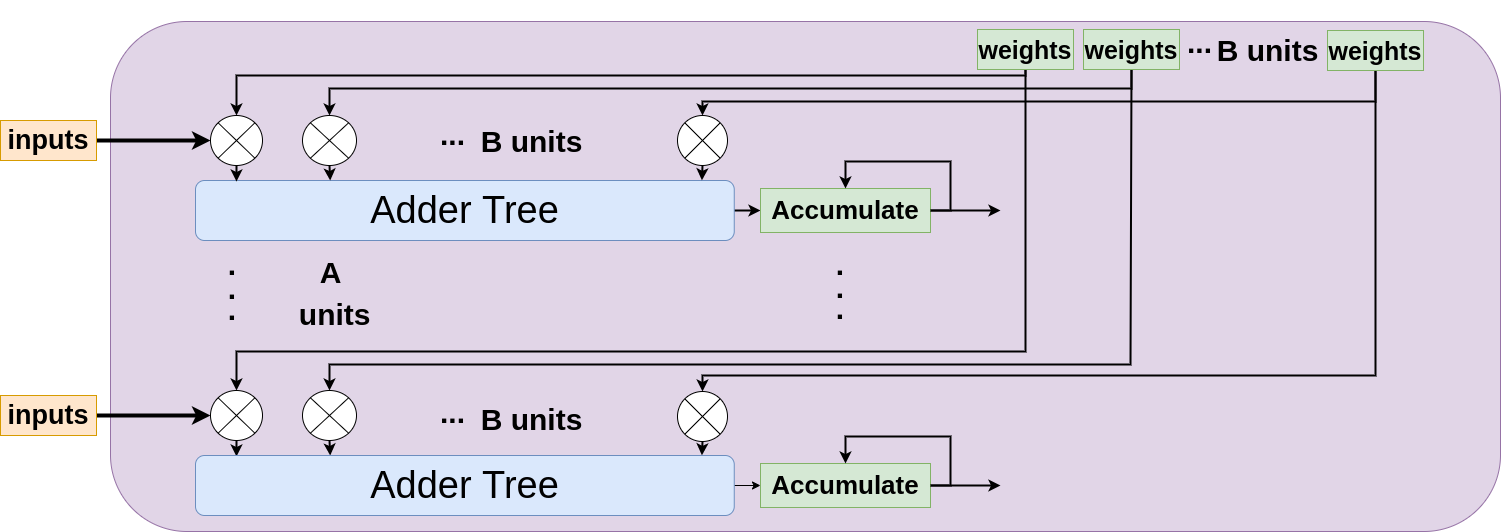} 
      \caption{PE block architecture - the rows share weights. For PE Blocks - 1, 2 \& 3, $A{=}k_1$ \& $B{=}k_2$, and for PE Blocks 4 \& 5, $A{=}k_3$ \& $B{=}k_4$}
      \label{label}
    \end{subfigure} \hspace{0.5cm}
    \caption{ViTA : Proposed design for the hardware accelerator}
    \label{fig:architecture}
        \vspace{-0.4cm}
\end{figure}

\begin{figure}[!t]
      \centering
      \includegraphics[width = 5cm]{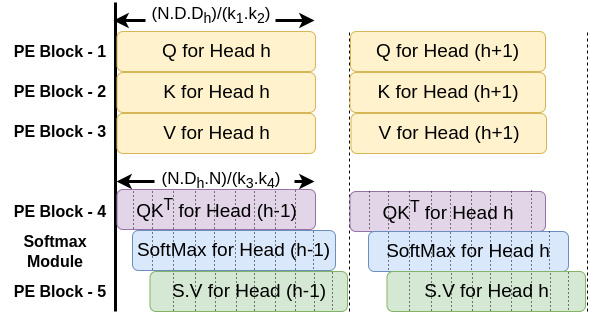}
      \caption{Scheduling the MSA computations. PE Block 4, Softmax Module, and PE Block 5 process in a row-granular pipeline fashion.}
      \label{fig:scheduling}
      \vspace{-0.4cm}
    \end{figure}

Drawing conclusions from the observations and implications made on accelerating the MLP and MSA layer, we propose a configurable processing element array-based hardware accelerator and the corresponding scheduling scheme, illustrated in Figs. \ref{fig:architecture} and \ref{fig:scheduling}. The PE blocks 1, 2, and 3 together make up the first compute engine that performs the computations for generating $Q$, $K$, and $V$, while the PE blocks 4 and 5 form the second compute engine that performs the $QK^T$ and $S.V$ operation, respectively. Dedicated units for SoftMax, LayerNorm, skip connections and non-linear activations have been included, with the former adapted from \cite{Lu2020}. We reuse the same PE blocks for computing the MSA concatenation and the MLP block results. The condition required for efficient acceleration of the MLP layer, as mentioned in Sec. \ref{sssec:mlp}, is to have equal MAC units dedicated to computing the hidden layer and output layer. This is realised by using half the rows in the PE blocks for computing the hidden layer and the other half for the output layer. 

As shown in Fig. \ref{fig:architecture}, $k_1\times{k_2}$ refers to the configuration of  PE blocks 1, 2 \& 3, while $k_3\times{k_4}$ refers to the configuration of PE block 4 \& 5, which together provide the configurability in our architecture. As illustrated in Fig. \ref{fig:scheduling}, in order to time match the MSA computations in the two compute engines and enable head-wise pipelining proposed in Sec. \ref{sssec:msa}, the optimal values of $k_1$, $k_2$, $k_3$, and $k_4$ should satisfy:
\begin{equation}
\frac{D}{k_1.k_2} = \frac{N}{k_3.k_4}    
\label{eq5}
\end{equation}
while maximising resource utilization subject to the on-chip resource constraints. Consequently, while PE Blocks 1, 2, and 3 operate on head $h$, PE Blocks 4 and 5 operate on head $h-1$. Within a head, PE Block 4, Softmax Module and PE Block 5 operate at a row granularity. 


\section{Analysis \& Experiments}\label{sec:analysis}

We evaluate the performance of ViTA across different model architectures in terms of the 
hardware utilization efficiency (HUE) \cite{hue} of the utilized resources on the FPGA. We choose a design configuration that works best for the ViT-B/16 model, while also restricting ourselves to the resources available on the Zynq ZC7020 MPSoC. From Eq. \ref{eq5}, this corresponds to the PE block configurations being $k_1 = 16$, $k_2 = 6$, $k_3 = 8$ \& $k_4 = 4$. We rely on the LUTs, rather than DSP slices, for compute capability, and ensure that all the memory can be held and accessed as intended from the BRAM cells. 
Although the DSP slices are power efficient, they are limited in number for the considered FPGA. Moreover, these are optimized for $18\textrm{bits}\times27\textrm{bits}$ operations, and performing $8\textrm{bits}\times8\textrm{bits}$ operations on these would lead to significant under-utilization. 

Although the design parameters of the PE array can be configured at run time, 
we demonstrate how having a single fixed configuration also allows for a run-time selection of the desired model and input image size, without a significant drop in efficiency. Since the other vision transformer models are typically made up of the MSA and MLP blocks, we can infer any of these models on ViTA by changing the control logic appropriately. For the case of Swin transformer, the W-MSA is performed on windows of $M\times{M}$, where $M{=}7$ by default. On the same hardware architecture, this would just correspond to the regular MSA being performed on $N{=}49$ repeatedly over these windows. Due to space constraints, we have not elaborated implementation aspects of other models in further detail.  

When synthesized on the Zynq ZC7020 MPSoC, the design is run at a clock of 150 MHz, and consumes a power of 0.88W for the given configuration. The obtained values of HUE and energy for processing an image are summarized in Table \ref{tab:hue}. We ensure that the latency in accessing the off-chip memory is hidden in all intermediate cases, as mentioned in Sec \ref{ss:had}, with the DRAM access bandwidth well under 1 word/cycle - a reasonable number for the considered FPGA. 
\begin{table}[h]
    \centering
    \begin{tabular}{l | l | l | l | l}
    \hfil Model & \hfil Image Dim. & \hfil HUE & \hfil fps & Energy (J)\\
    \hline
    \hfil ViT-B/16 & $256\times256\times3$  & \hfil 93.2\% & \hfil 2.17 & \hfil 0.406 \\
    \hfil ViT-B/16 or DeiT-B & \hfil $224\times224\times3$  & \hfil 92.8\% & \hfil 2.75 & \hfil 0.320\\ 
    \hfil DeiT-S & \hfil $224\times224\times3$  & \hfil 87.2\% & \hfil 9.36 & \hfil 0.094\\ 
    \hfil DeiT-T & \hfil $224\times224\times3$  & \hfil 66.2\% & \hfil 19.01 & \hfil 0.046 \\
    \hfil Swin-T   & \hfil $224\times224\times3$  & \hfil 81\% & \hfil 8.71 & \hfil 0.101 \\
    \end{tabular}
    \caption{Overall hardware utilization efficiency (HUE), Frame Rate (fps) and energy for processing one image for different model architectures - for the optimal architecture configuration chosen for ViT-B/16 with a 256$\times$256$\times$3 image dimensions}
    \label{tab:hue}
        \vspace{-0.2cm}
\end{table}

Table \ref{tab:trans-com} compares the performance our design against other vision transformer accelerators. As illustrated, ViTA achieves a significant reduction in power consumption.
Since ViTA is focused on highly resource-constrained environments, it is difficult to compare it against other designs in terms of the processing frame rate. However, for the lower compute capability and technology node of our target device, the performance scales comparably with other works. Furthermore, the frame rates achieved by ViTA for the smaller model variants, such as DeiT-S, DeiT-T, and Swin-T are reasonable for embedded applications, such as drone navigation. Moreover, an FPGA environment makes ViTA reconfigurable for different model configurations for improved performance. This coupled with the energy efficiency of ViTA makes it a good choice for deployment on the edge.
\begin{table}[h]
    \centering
    \begin{tabular}{l | l | l |l |l}
    \hfil Accelerator Design & \hfil Target Device & \hfil Power (W) & \hfil fps & \hfil fps/W\\
    \hline
    \hfil Row-wise-acc. \cite{vit-hw} & \hfil ASIC (40nm) & \hfil * & \hfil 44.5 & \hfil * \\
    \hfil Auto-vit-acc. \cite{auto-vit} & \hfil FPGA (16nm)  & \hfil 9.40 & \hfil 25.9 & \hfil 2.76\\ 
    \hfil \textbf{ViTA (ours)} & \hfil FPGA (28nm) & \hfil \textbf{0.88} & \hfil 2.75 & \hfil 3.12\\ 
    \end{tabular}
    
\hfill * not reported
    \caption{Performance comparison of vision transformer accelerators for DeiT-B on $224\times224\times3$ dimensioned images}
    \label{tab:trans-com}
        \vspace{-0.4cm}
\end{table}

\section{Conclusions}\label{sec:conclusions}
We proposed ViTA - a configurable hardware accelerator and dataflow design that can be employed for inference of ViT models on resource-constrained edge devices. In particular, we introduced a head-level coarse-grained pipeline and performed inter-layer optimization on the MLP layer to avoid unnecessary intermediate result staging. Our design avoids repeated off-chip memory accesses, achieves a high resource utilization efficiency of about 90\%, and reports a significantly low power of 0.88W while having a reasonable frame rate - making it well suited for various edge applications.

\bibliographystyle{IEEEtran}
\bibliography{IEEEabrv,bibliography}


\end{document}